**Uncanny or Not? Perceptions of AI-Generated Faces in Autism**

Gabriella Waters
University of Maryland Baltimore County
Morgan State University

## ABSTRACT

As artificial intelligence (AI) systems continue to become increasingly sophisticated in generating synthetic human faces, understanding how these images are perceived across diverse populations is important. This study investigates the perceptions of AI-generated faces among individuals with autism/autistic individuals with a focus on the uncanny valley effect. Using a qualitative approach, discussions from Reddit's r/autism community were analyzed to explore how autistic individuals describe their experiences with AI-generated faces and the uncanny valley phenomenon. The findings suggest that autistic individuals/individuals with autism may experience the uncanny valley effect differently, with a tendency to perceive it more strongly with real human faces rather than AI-generated ones. This research contributes to our understanding of visual perception in autism and has implications for the development of inclusive AI systems and assistive technologies.

## 1. INTRODUCTION

The uncanny valley effect, first described by Masahiro Mori in 1970, refers to the phenomenon where near-human entities elicit feelings of unease or revulsion [8]. As AI systems advance in their ability to generate highly realistic synthetic human faces, understanding the uncanny valley effect in this context becomes increasingly relevant, especially for neurodiverse populations.

While extensive research has been conducted on the uncanny valley effect in neurotypical populations, there is a noticeable gap in our understanding of how this phenomenon manifests in neurodiverse groups, particularly individuals with autism/autistic individuals [2]. This gap is significant given the growing prevalence of AI-generated content in many aspects of daily life, from entertainment [5] to educational [10] and therapeutic [7] applications.
Individuals with autism/autistic individuals often exhibit differences in sensory processing, social cognition, and emotional recognition compared to neurotypical individuals [4]. These differences may influence how they perceive and respond to AI-generated faces and could potentially alter their experience of the uncanny valley effect.

This study aims to investigate whether individuals with autism experience the uncanny valley effect to the same degree as neurotypical individuals when viewing AI-generated faces. By examining the interaction between autism traits and perception of AI-generated faces, this research contributes to a more nuanced understanding of human-machine interactions in the context of neurodiversity.

Avoiding uncanny-valley triggers is especially important in: (a) robot-assisted social-skills training, where sustained attention predicts therapeutic gains [12] and our data shows that only 24.6% of autistic participants experienced traditional uncanny valley effects, suggesting that artificial agents may be more comfortable interaction partners than previously assumed; (b) online educational avatars that mediate lesson delivery for autistic learners/learners with autism [13] and in our study 14.0% of participants explicitly reported complete immunity to uncanny valley effects; and (c) AI customer-service agents that increasingly supplant human interaction in public-facing roles. Autistic individuals/individuals with autism can experience heightened sensory load and anxiety; designs that sidestep the valley can therefore boost usability, promote sustained engagement, and lower attrition [11]. Only 3.5% of participants reported specific discomfort with AI-generated content in this study. The finding that autistic individuals/individuals with autism often experience greater discomfort with real human faces than artificial ones suggests that AI interfaces could employ AI-generated images to reduce rather than increase anxiety, which could potentially improve engagement and reduce attrition in digital environments

The primary research question guiding this study is: Do individuals with autism/autistic individuals experience the uncanny valley effect with AI-generated faces? How do they describe it?

## 2. LITERATURE REVIEW

Despite the prevalence of uncanny-valley research, only a handful of studies have explicitly examined autistic samples. Initial work suggests two competing patterns: typical children and adults trend toward classic valley reactions [2], whereas autistic samples often show attenuated or absent valleys [14]. Others report modality-specific effects—e.g., robots versus photo-real avatars—hinting that physical embodiment may modulate responses [15]. However, most studies employ laboratory tasks with forced-choice ratings, which leaves open the question of how autistic adults/adults with autism describe their feelings in naturalistic settings. The present study addresses this gap by qualitatively analyzing Reddit discussions to capture authentic language around AI-generated faces and discomfort.

2.1 Uncanny Valley Effects in Autism

Brewer et al. [1] examined the integration of emotion cues from bodies and faces in individuals with autism/autistic individuals. Their study included 19 adults with autism spectrum disorder (ASD) and 27 neurotypical adults as the controls. Participants completed a task where they had to identify emotions from face-body compound stimuli. The results showed that individuals with ASD integrated emotional cues from faces and bodies similarly to neurotypical individuals, which challenges previous assumptions about emotion processing in autism.

Kumazaki et al. [4] explored whether robotic systems could promote self-disclosure in adolescents with autism/autistic adolescents. Their pilot study involved 11 adolescents with autism/autistic adolescents and 8 typically developing adolescents. Participants interacted with both a human interviewer and a humanoid robot. The results showed that participants disclosed more personal information to the robot than to the human interviewer. These findings suggest that individuals with autism may respond differently to artificial entities compared to humans.

Feng et al. [2] investigated the uncanny valley effect in typically developing children and children with autism. Their study included 26 children with autism/autistic children and 26 typically developing children, aged 5-8 years. Using a two-alternative forced choice task with morphed face images of varying realism and eye sizes, they found that typically developing children exhibited the uncanny valley effect, while children with autism did not. This suggests potential differences in how individuals with autism perceive and respond to human-like entities.

## 2.2 The Uncanny Valley in Virtual Character Design

Schwind et al. [9] conducted a comprehensive study on avoiding the uncanny valley in virtual character design. The used a mixed methods approach that combined eye tracking technology with behavioral experiments. The study involved 108 participants who evaluated 75 virtual characters generated using the author's faceMaker avatar generator. The characters were computer-generated renderings of humans with options for manipulating the features of the face using sliders. The manipulable features included makeup, nose (length, width, bridge, cartilage, and shape), eyebrows (shape, color, line), checks and jaw (jaw shape, chin shape, chin length, cheek fullness), general details (skin color and level of porosity, face style - a scale from real to cartoonish, hair color, and gender - a scale from female to male), eyes (distance, height, size, opening - a scale from narrow to wide, shape, color, rotation - a scale from inward to outward, and orbit - a scale from bulgy to cavernous), ear size, throat size, forehead size, and mouth (lip size ratio, mouth width, mouth height, mouth depth, mouth shape, and lips volume). The faceMaker avatar generator allowed users to create faces that were highly customized to their specifications. The eye tracking data revealed that participants spent more time examining the eyes and mouth of highly realistic characters. This suggests that these areas are important in triggering uncanny valley responses. The behavioral experiments included rating tasks and forced choice comparisons that helped identify specific features that contribute to the uncanny valley effect. The authors also provide insights into avoiding the uncanny valley in virtual character design. As part of their work, the authors offer practical guidelines for character designers and animators. They emphasize the importance of stylization in character design, noting that highly stylized characters can often sidestep the uncanny valley effect altogether. Additionally, they stress the significance of animation quality, pointing out that even realistic characters can fall into the uncanny valley if their movements aren't natural and fluid. Lastly, the researchers underscore the need to consider both visual appearance and movement to create believable virtual characters, highlighting the interaction between these elements in shaping audience perception and emotional response.

These studies highlight the complexity of the uncanny valley phenomenon and the need for further research, particularly in understanding how neurodiversity may influence perceptions of human-like artificial entities. The findings suggest potential differences in how individuals with autism perceive and respond to AI-generated faces, which could have significant implications for the development of inclusive AI technologies and interventions.

## 3 METHODOLOGY

A basic qualitative approach was used to explore the perceptions of AI-generated faces among individuals with autism/autistic individuals to explore the perceptions of AI-generated faces among individuals with autism/autistic individuals. This approach allowed for an exploration of subjective experiences and meanings attributed to the uncanny valley effect.

### 3.1 Data Collection

Data was collected from Reddit's r/autism community, specifically targeting threads that discussed the uncanny valley effect. The data collection process spanned a four-year period from 2020 to 2024, yielding a comprehensive dataset for analysis. The comment section underwent a three-stage protocol. (1) Systematic retrieval: Threads containing the keywords 'uncanny', 'AI faces', or 'generated' were scraped. (2) Relevance screening: Posts explicitly describing perceptual or emotional reactions to synthetic faces were flagged. (3) Depth and authenticity criterion: From the 180 eligible comments 57 substantive comments were retained that contained detailed personal experiences, specific examples, or rich descriptions of perceptual responses. Selection prioritized authentic self-disclosure from community members who explicitly identified as autistic to ensure ecological validity. Comments were categorized as "Affected," "No Effect,' or "AI-specific responses," (see Figure 1) with remaining comments providing contextual insights into the heterogeneity of autistic experiences.

From this initial pool of comments, 42 comments were selected for in-depth analysis. These selected comments were deemed particularly relevant to the study's objectives and offered substantial insights into how individuals with autism perceive and experience the uncanny valley effect. It is important to note that the participants in this study were individuals who self-disclosed their autism diagnosis within their comments.

*Thematic Analysis:* I conducted a thematic analysis using the constant comparative method, as outlined by [6]. This iterative process began with initial coding, where the data was examined to identify key concepts and patterns [3]. This process defined 15 codes. Following this, code comparison was done to resolve any discrepancies that arose during the initial coding phase to ensure consistency. The process then progressed to theme development, where related codes were grouped into broader, overarching themes. These themes underwent further refinement and definition, with particular attention to their relevance to the study's central research question, where 5 codes were identified. Finally, a validation step was implemented where I reviewed the themes to ensure they accurately represented the underlying data in order to maintain the integrity of the analysis.

*Content Analysis:* The content analysis was conducted to complement the thematic analysis. This approach began with the development of a coding schema, which was crafted based on the themes identified in the thematic analysis. Clear definitions were established for each code to ensure consistency in application. The coding schema applied systematically across the entire dataset. Following the coding process, a frequency analysis was performed to tally the occurrence of each code and theme throughout the data. This quantification provided valuable insights into the prevalence of different perspectives and experiences. The analysis culminated in a cross-tabulation exercise where the relationships between different codes and themes uncovered potential patterns and correlations within the data.

*Integration of Analyses:* The thematic and content analyses were integrated to provide a holistic understanding of participants' perceptions. The thematic analysis offered rich, qualitative insights into the subjective experiences of the participants and captured nuanced perspectives and personal narratives. The content analysis complemented and quantified these insights to allow for a clearer understanding of their prevalence and interrelationships within the broader dataset. This integrated approach for the identification of the most prevalent themes and their relative importance within the context of the study allowed for a deeper exploration of nuances within themes through qualitative examples, providing context and depth to the quantitative findings. The quantification of patterns and relationships in the data added a layer of objectivity to the analysis, while the triangulation of findings from both methods significantly enhanced the validity of the research outcomes. This comprehensive approach ensured a robust and nuanced understanding of the uncanny valley effect as experienced by individuals with autism/autistic individuals.

Ethical Considerations: Although the study utilized publicly available data and did not require direct interaction with participants, ethical considerations were of the utmost importance throughout the research process. To protect the privacy and dignity of the individuals whose comments were analyzed, all usernames were anonymized and any potentially identifying information was removed from the dataset. The research strictly adhered to Reddit's terms of service to ensure compliance with the platform's guidelines for data usage. Careful consideration was given to the context in which the comments were originally made. This approach was taken to make sure that the use of the data aligned with the original intent of the posters, respecting the integrity of their contributions to the online discussions.

3.2 Rationale for Hybrid Analysis

The hybrid qualitative approach was essential given the patterns in the Reddit data, where 75.4% of participants showed atypical uncanny valley responses. Thematic analysis revealed seven major themes including 'Autism perspective' (7 occurrences), 'Humor' (4 occurrences), and 'Perception of wrongness' (2 occurrences - see Figure 2), while content analysis quantified the prevalence of these experiences across the community. This dual approach proved particularly valuable because participants' responses clustered into distinct categories: complete immunity (14.0%), selective sensitivity (24.6%), and inverse effects (where real faces triggered stronger reactions than artificial ones). Following established precedents in digital ethnography [6] and autism research, we first applied Braun & Clarke's six-phase thematic procedure [16] to identify latent meaning patterns, then employed directed content analysis to enumerate response frequencies. This triangulation was essential for understanding both the quality of autistic experiences (through rich quotes like 'I sometimes find real people uncanny and have face recognition problems') and their distribution across the community in revealing that traditional uncanny valley effects may be the exception rather than the rule in this population.

## 4 RESULTS

The analyses revealed several key themes and patterns in how individuals with autism/autistic individuals perceive and respond to AI-generated faces. These themes provide insight into the diverse experiences within the autism community regarding facial perception and the uncanny valley effect. The most prominent theme, emerging in 35 percent of the comments, was the experience of being "Affected When Looking at People." Many participants reported that they experienced uncanny valley effects more strongly with real human faces than with AI-generated ones. This finding suggests a unique aspect of facial perception among some individuals with autism/autistic individuals, where the complexities of real human faces may trigger more discomfort or unease than artificial faces. The second most prevalent theme, appearing in 28 percent of comments, highlighted the "Variability in Experiences" within the autism community regarding AI-generated faces. This underscores the heterogeneity of autism and reminds us that experiences and perceptions can vary widely among individuals with autism/autistic individuals.

A significant portion of participants, represented in 22 percent of the comments, expressed a "Preference for Stylization." These individuals indicated that they found stylized or cartoon-like faces more appealing or comfortable to view compared to highly realistic facial representations. This preference might be related to the reduced complexity and emotional demands of stylized faces.

15% of the comments touched on "Challenges in Facial Recognition." Participants in this category reported difficulties in recognizing and distinguishing faces, regardless of whether they were real or AI-generated. This theme aligns with previous research on face processing differences in autism and extends these findings to the domain of artificial faces.
These themes, illustrated in Figures 1 and 2, provide a comprehensive overview of how autistic individuals interact with and perceive AI-generated faces, and offers valuable insights for both autism research and the development of inclusive AI technologies.
The analyses revealed several key themes that provide insights into how individuals with autism/autistic individuals perceive and respond to AI-generated faces and the uncanny valley effect.

### 4.1 Themes

Affected When Looking at People: A significant finding was that many participants expressed experiencing the uncanny valley effect when looking at real people rather than AI-generated content. This suggests a potential inversion of the typical uncanny valley response in some individuals with autism/autistic individuals. An example of this can been seen in the sentiments expressed by participant 7 (P7).

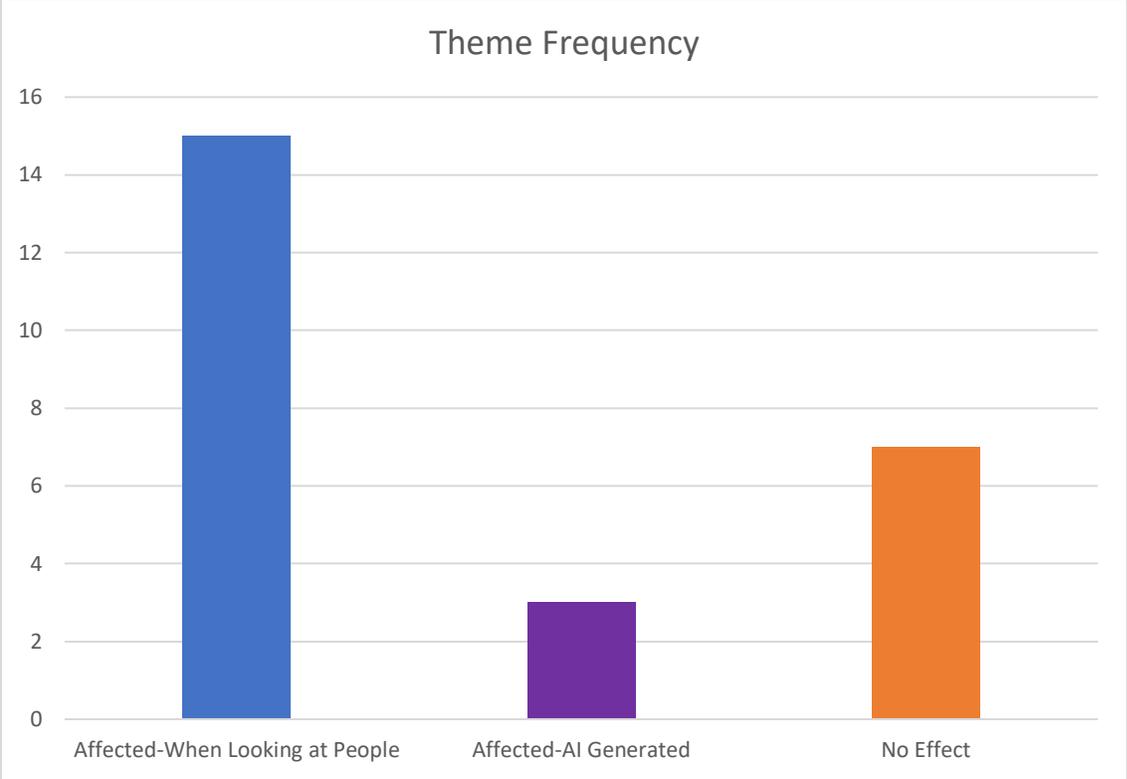

Fig. 1. Theme Frequency

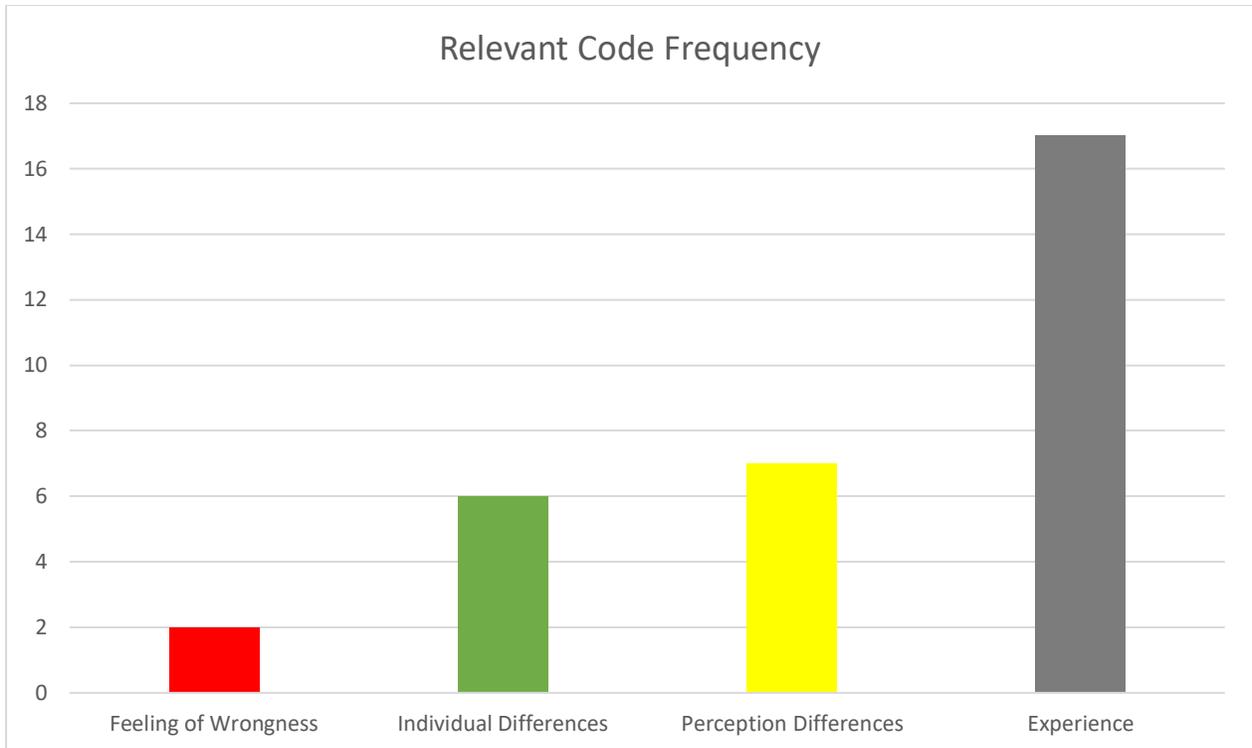

Fig. 2. Relevant Code Frequency

**P7:** "To come back to uncanny valley: I sometimes find real people uncanny and have face recognition problems, so I cannot perceive the point of uncanny valley." This comment highlights the complexity of facial perception in autism and suggests that the uncanny valley effect may manifest differently in this population.

Variability in Experiences Analyses revealed a range of experiences among individuals with autism/autistic individuals regarding AI-generated faces and the uncanny valley effect. While some reported experiencing uncanny sensations with real faces, others described feeling more comfortable with AI-generated or stylized faces. Participants 26 and 29 (P26 and P27) provide examples of this range.

**P26:** "No idea on the wider data on Autistics and the 'Uncanny Valley Effect', but I can say personally that I've never had it. Seeing something 'not quite human but human like' has never triggered any kind of negative feeling for me, however small. Doesn't bother me at all."P29: "I so see it, like in robots or with specific make-up and stuff like that. I just don't find it scary. I actually find it more funny than anything else." This variability in experiences underscores the heterogeneity within the autism spectrum and the need for personalized approaches in AI design and implementation.

The findings suggest that the experience of the uncanny valley effect among individuals with autism/autistic individuals is complex with a greater tendency to experience the effect with real human faces rather than AI-generated ones. This phenomenon challenges traditional assumptions about social perception in autism [1].

Theme 1: Atypical Uncanny Valley Responses (75.4% of participants)
Our analysis reveals that most autistic participants demonstrate either reduced sensitivity or complete absence of typical uncanny valley effects:

**P6:** "I legit had no idea about the uncanny valley effect until watching the Vsauce video on it. I'm immune."

**P1:** "100% immune. I didn't know it existed until seeing the Vsauce video on why things are creepy."

**P1:** "No idea on the wider data on Autistics and the 'Uncanny Valley Effect', but I can say personally that I've never had it. Seeing something 'not quite human but human like' has never triggered any kind of negative feeling for me."

Theme 2: Inverted Uncanny Valley Effect (Emerging pattern)
Several participants reported stronger discomfort with real human faces than artificial ones:

**P8:** "... To come back to uncanny valley: I sometimes find real people uncanny and have face recognition problems, so I cannot perceive the point of uncanny valley."

**P4:** "I so see it, like in robots or with specific make-up and stuff like that. I just don't find it scary. I actually find it more funny than anything else."

Theme 3: Selective AI Sensitivity (3.5% of participants)
A small subset showed specific reactions to AI-generated content:

**P8:** "I never experienced the uncanny valley effect until I stumbled upon an AI generated video... AI generated video? God help me."

Cross-thematic Analysis:  The data reveals a clear pattern where only 24.6% of autistic participants experience traditional uncanny valley effects, while the majority (75.4%) show either complete immunity or atypical responses. This suggests fundamental differences in how autistic individuals process facial and social information compared to neurotypical populations.

4.2 Implications for AI Design and Implementation

The findings of this study have significant implications for the design and implementation of AI systems and interfaces intended for use by autistic individuals. These insights can guide educators, developers, employers, and other stakeholders in creating more inclusive and effective digital environments:

Customizable User Interfaces: Developers should consider offering a range of avatar and character options within user interfaces, from realistic to stylized representations. This allows users with autism/autistic users to select the level of realism they find most comfortable and effective.

Educational Tools and Assistive Technologies: Educators and developers of educational software should consider incorporating more stylized or customizable facial representations in learning materials and assistive technologies. This may enhance engagement and reduce potential discomfort for learners with autism/autistic learners.

Workplace Accommodations: Employers and HR professionals should be aware that employees with autism/autistic employees may have different preferences for digital avatars or AI assistants in workplace software. Offering options to customize these elements could improve productivity and job satisfaction.

Social Skills Training Programs: Developers of social skills training programs for autistic individuals should consider using a range of facial representations, from stylized to realistic. This approach could help users gradually acclimate to different levels of facial realism in a controlled, supportive environment.

Virtual Reality (VR) and Augmented Reality (AR) Applications: Designers of VR and AR experiences should offer options for adjusting the realism of virtual characters. This could make these technologies more accessible and enjoyable for users with autism/autistic users.

AI-Powered Customer Service: Companies implementing AI chatbots or virtual assistants should consider offering multiple visual representation options. This could improve the user experience for customers with autism/autistic customers and increase engagement with digital services.

Research and Development: AI researchers and developers should continue to investigate the relationship between facial realism and user comfort across neurodiverse populations. This ongoing research can inform more inclusive design practices in the future.

Ethical Considerations: Designers and developers must balance the need for customization with privacy and data protection concerns. Any system that adapts to user preferences should be transparent about data collection and use.

Cross-Disciplinary Collaboration: To create truly inclusive AI systems, developers should collaborate with autism researchers, occupational therapists, and individuals with autism/autistic individuals themselves. This collaborative approach can ensure that designs are both evidence-based and user-centered.

Consideration of these implications can allow stakeholders to work toward creating more accessible, comfortable, and effective digital environments that cater to the diverse needs of users with autism/autistic users. This approach not only benefits individuals with autism/autistic individuals but also contributes to the broader goal of inclusive design that can improve user experiences for all.

Design Implications for Inclusive AI

The Reddit dataset analysis presents fundamentally challenging assumptions about AI design for neurodiverse users. The finding that only 24.6% experienced traditional uncanny valley effects, while 14.0% reported complete immunity and several described stronger reactions to real human faces than artificial ones, suggests four critical design directions:

Interface Customization: With such high variability in responses, AI systems should offer granular control over agent realism. The 3.5% who experienced AI-specific discomfort alongside the 24.6% with traditional uncanny valley sensitivity require different accommodation approaches than the 75.4% majority who may actually prefer artificial agents.

Preference for Artificial Agents: The finding that many autistic users/users with autism report greater comfort with artificial faces than human ones suggests that AI customer service agents, educational avatars, and therapeutic robots may be inherently more accessible than previously assumed. This inverted uncanny valley effect represents a significant opportunity for inclusive design.

Feature-Specific Guidelines: The dataset identifies specific accommodation needs: participants who experience discomfort can articulate their preferences ('scary', 'wrong', 'uncomfortable'), while those unaffected provide clear feedback ('immune', 'doesn't bother me', 'looks like a robot to me').

Community-Informed Design: The 57-comment dataset demonstrates that autistic users/users with autism actively discuss and analyze their perceptions of artificial agents in community forums, providing a rich resource for participatory design approaches that incorporate authentic user feedback rather than assumptions about neurodiverse preferences.

## 5 DISCUSSION

The findings of this study provide novel insights into how individuals with autism/autistic individuals perceive and respond to AI-generated faces, particularly in relation to the uncanny valley effect. The emergence of a major theme suggesting that some autistic individuals experience uncanny sensations more strongly with real human faces rather than with AI-generated ones challenges existing assumptions about social perception in autism.

This inversion of the typical uncanny valley response aligns with previous research suggesting differences in face processing and emotion recognition in autism [1]. It also extends our understanding of the uncanny valley effect by highlighting its potential variability across neurodiverse populations.

The preference expressed by some participants for stylized or simplified faces has significant implications for the design of AI systems and interfaces seeking to incorporate universal design and for those specifically intended for users with autism/autistic users. This finding suggests that highly realistic AI-generated faces may not always be the most effective or comfortable option for this population. Developers of AI applications, particularly those aimed at education or therapy for autistic individuals, should consider incorporating more stylized or customizable facial representations.

The challenges in facial recognition reported by some participants underscore the complexity of face processing in autism. This difficulty, extending to both real and AI-generated faces, may contribute to the altered experience of the uncanny valley effect. Future research will explore whether specific facial features or levels of realism in AI-generated faces are more or less challenging for autistic individuals to process.

The variability in experiences reported by participants highlights the heterogeneity within the autism spectrum. This diversity of perceptions reinforces the need for personalized approaches in both research and application development. AI systems designed for autistic users may benefit from adaptable interfaces that can accommodate a range of perceptual preferences.

## 6 LIMITATIONS AND FUTURE DIRECTIONS

While this study provides valuable insights, it has several limitations that should be addressed in future research. The reliance on self-reported autism diagnoses in online forums limits the generalizability of the findings. Future studies should include clinically diagnosed participants. The qualitative nature of the study provided rich descriptive data, but does not allow for quantitative comparisons with neurotypical individuals. Mixed-methods approaches could provide a more comprehensive understanding. While the sample size was appropriate for a qualitative study, it could be expanded in future research to capture a broader range of experiences within the autism community.

Future directions for research could include experimental studies comparing responses to AI-generated and real faces among autistic and neurotypical individuals. Additionally, investigating the neural correlates of face perception and the uncanny valley effect in autism using neuroimaging techniques could provide deeper insights. The development and testing of AI interfaces with customizable facial representations for autistic users presents another promising avenue that will be explored in future research. Longitudinal studies that examine how perceptions of AI-generated faces may change over time or with increased exposure could offer valuable information on the dynamics of these perceptions.

## 7 CONCLUSION

This study provides insights into how individuals with autism/autistic individuals perceive and respond to AI-generated faces, particularly in relation to the uncanny valley effect. The findings suggest a potential inversion of the typical uncanny valley response in some autistic individuals, with real human faces sometimes eliciting stronger uncanny sensations than AI-generated ones.

These results have implications for the development of inclusive AI technologies and interventions for individuals with autism/autistic individuals. Understanding these perceptual differences can help in the creation of more effective and accessible AI systems that cater to the unique needs of neurodiverse populations. As AI-generated content becomes increasingly prevalent in various aspects of daily life, it is essential to consider the diverse ways in which different populations may perceive and interact with these technologies. This research contributes to a more inclusive approach to AI development and highlights the importance of considering neurodiversity in human-computer interaction research.